\documentclass[twocolumn,aps,pra,reprint,superscriptaddress,longbibliography]{revtex4-1}

\usepackage{amsmath}
\usepackage{graphicx}
\usepackage[lofdepth,lotdepth, caption=false]{subfig}
\usepackage{verbatim}
\usepackage{color}
\usepackage{dcolumn}
\usepackage{bm}
\usepackage{miller}
\usepackage{color}
\usepackage{xr-hyper}
\usepackage{hyperref}
\usepackage[english]{babel}

\begin{document}

\title{T$_{d}$ to 1T$^{\prime}$ structural phase transition in WTe$_{2}$ Weyl semimetal}

\author{Yu Tao}
\affiliation{Department of Physics, University of Virginia, Charlottesville,
Virginia 22904, USA}

\author{John A.~Schneeloch}
\affiliation{Department of Physics, University of Virginia, Charlottesville,
Virginia 22904, USA}

\author{Adam A.~Aczel}
\affiliation{Neutron Scattering Division, Oak Ridge National Laboratory, Oak Ridge, Tennessee 37831, USA}
\affiliation{Department of Physics and Astronomy, University of Tennessee, Knoxville, Tennessee 37996, USA}

\author{Despina Louca}
\thanks{Corresponding author}
\email{louca@virginia.edu}
\affiliation{Department of Physics, University of Virginia, Charlottesville,
Virginia 22904, USA}

\begin{abstract}

Elastic neutron scattering on a single crystal combined with powder X-ray diffraction measurements were carried out to investigate how the crystal structure evolves as a function of temperature in the Weyl semimetal WTe$_{2}$. A sharp transition from the low-temperature orthorhombic phase (T$_{d}$) to the high-temperature monoclinic phase (1T$^{\prime}$) was observed at ambient pressure in the single crystal near $\sim$565 K. Unlike in MoTe$_{2}$, the solid-solid transition from T$_{d}$ to 1T$^{\prime}$ occurs without the cell doubling of the intermediate T$_{d}^{*}$ phase with AABB (or ABBA) layer stacking. In powders however, the thermal transition from the T$_{d}$ to the 1T$^{\prime}$ phase was broadened and a two phase coexistence was observed until 700K, well above the structural transition. 

\end{abstract}
\maketitle

\section{Introduction}

Transition metal dichalcogenides (TMDs) have attracted considerable attention recently because of their intriguing electronic band structure properties that render them hosts to exotic quasiparticles. MoTe$_{2}$ and WTe$_{2}$ are reported to be type-II Weyl semimetals in the orthorhombic T$_{d}$ phase \cite{deng_experimental_2016, wu_observation_2016} due to spatial inversion symmetry breaking, and both show a large non-saturating magnetoresistance  \cite{yang_elastic_2017, ali_large_2014, cai_drastic_2015}. 
They are layered structures, held together by van der Waals forces, and can undergo multiple solid-solid transitions through the sliding of layers \cite{clarke_low-temperature_1978,brown_crystal_1966}. Upon quenching from high temperatures, the monoclinic phase, 1T$^{\prime}$, was first shown to be stabilized in MoTe$_{2}$, from which the low-temperature orthorhombic phase (T$_{d}$) emerges. The high-temperature monoclinic phase \cite{clarke_low-temperature_1978} and the low-temperature orthorhombic phase differ in their layer stacking. In WTe$_{2}$, on the other hand, only the T$_{d}$ phase has been reported at ambient pressure, and the 1T$^{\prime}$ phase has been theoretically predicted to be absent up to at least 500 K\cite{kim_origins_2017}. Application of external pressure, however, leads to a T$_{d}$ to 1T$^{\prime}$ phase transition that commences around 6.0 GPa \cite{zhou_pressure-induced_2016}. 

The 1T$^{\prime}$ crystal structure is shown in Fig.\ \ref{fig:fig1}(a), projected in the $a$-$c$ plane. Layer stacking follows two possible ordering schemes, with stacking types labeled ``A'' and ``B'' (Fig.\ \ref{fig:fig1}(b)) \cite{schneeloch_emergence_2019,tao_appearance_2019}. The T$_{d}$ phase is constructed by stacking either AAAA...\ or BBBB...\ sequences, while the 1T$^{\prime}$ is built by stacking ABAB...\ or BABA...\ layers. We recently reported that an intermediate pseudo-orthorhombic T$_{d}^{*}$ phase appears across the transition boundary between T$_{d}$ and 1T$^{\prime}$, with an AABB...\ (or ABBA...) layer stacking in MoTe$_{2}$. The T$_{d}^{*}$ phase is only observed upon warming, while on cooling, diffuse scattering is seen, most likely arising from a frustrated tendency towards the T$_{d}^{*}$ stacking order. \cite{tao_appearance_2019, dissanayake_electronic_2019}. Regardless of A- or B-type stacking, all pairs of neighboring layers are positioned relative to each other in essentially the same way, which can be captured by an in-plane displacement parameter $\delta$ \cite{schneeloch_evolution_2020}, as shown in Fig.\ \ref{fig:fig1}(a). We define $\delta$ as the distance along the $a$-axis between the midpoints of metal-metal bonds of neighboring layers; this definition is uniquely defined for both 1T$^{\prime}$ (where it is related to the $\beta$ angle) and T$_{d}$.

With W substitution as in Mo$_{1-x}$W$_{x}$Te$_{2}$, the 1T$^{\prime}$ to T$_{d}$ structural transition temperature increases up until $x \approx 0.57$ \cite{yan_composition_2017}. However, it is not known at present whether this transition occurs at ambient pressure at the other end of the phase diagram with $x = 1$ as in WTe$_{2}$. A pressure-driven T$_{d}$--1T$^{\prime}$ structural transition has been reported to appear at 4 - 5 GPa \cite{lu_origin_2016}, at 8 GPa \cite{xia_pressure-induced_2017}, and in a broad range from 6.0 to 18.2 GPa, during which a volume collapse with dramatic changes in the lattice constants was observed \cite{zhou_pressure-induced_2016}. In MoTe$_{2}$, pressure suppresses the temperature of the 1T$^{\prime}$-T$_{d}$ transition, and extinguishes it by $\sim$1.2 GPa \cite{Qi2016,heikes_mechanical_2018,dissanayake_electronic_2019}, though dramatic changes in the lattice constants between the phases have not been reported. Nonetheless, the presence of a transition in WTe$_{2}$ under pressure, as well as the trend of increasing T$_{d}$--1T$^{\prime}$ transition temperature with W-substitution in the Mo$_{1-x}$W$_{x}$Te$_{2}$ \cite{oliver_structural_2017, yan_composition_2017, lv_composition_2017, rhodes_engineering_2017} phase diagram suggest the possibility of an ambient-pressure transition at high temperatures. 

Using elastic neutron scattering, we observed the T$_{d}$--1T$^{\prime}$ structural phase transition at ambient pressure in a single crystal of WTe$_{2}$. The transition is sharp, occurs at $\sim$565 K, and proceeds without hysteresis. No intermediate phase is present across the phase boundary in WTe$_{2}$, in contrast to the T$_{d}^{*}$ phase seen in MoTe$_{2}$. From powder X-ray diffraction (XRD) however, the transition appears broad and incomplete up to 700 K, with phase coexistence across a wide temperature range.

\section{Experimental Details}

The WTe$_{2}$ single crystals were grown out of a Te flux. First, WTe$_{2}$ powder was prepared from stoichiometric ratios of W and Te powders. The sintering was done in an evacuated quartz silica ampoule at 900 $^{\circ}$C for 2 days. The sintered powder was then pressed into a pellet and sealed with excess Te in a molar ratio of 1:13. The ampoule was placed horizontally in a tube furnace and heated at a constant temperature of 850 $^{\circ}$C for 7 days. Excess Te was removed by re-inserting one end of the ampoule into a tube furnace at $\sim$900 $^{\circ}$C and decanting the molten Te towards the cold end. For XRD, powder was sintered as described above.

Resistivity measurements under magnetic fields of 0 and 9 T are shown in Fig.\ \ref{fig:fig1}(c). The residual resistivity ratio (RRR) from the 0 T data is calculated to be $\sim$118(3). Our WTe$_{2}$ crystals also have a large magnetoresistance, with a magnitude of 51,553\% at 2 K under a 9 T magnetic field. 
These values are reasonably high \cite{lv_dramatically_2016}, though higher values have been reported in the literature, such as an RRR of $\sim$370 and a magnetoresistance of 452,700\% at 4.5 K in an applied field of 14.7 T \cite{ali_large_2014}. 

Elastic neutron scattering was performed on the triple axis spectrometer HB1A, located at the High Flux Isotope Reactor at Oak Ridge National Laboratory (ORNL). The elastic measurements used an incident neutron energy of 14.6 meV and the collimation was 40$^{\prime}$-40$^{\prime}$-S-40$^{\prime}$-80$^{\prime}$. The crystal was mounted to an aluminum plate via aluminum wire, and a furnace was used to control the temperature. Powder XRD measurements were collected as a function of temperature between 300 K and 700 K. %
Rietveld refinement was done using the \textsc{GSAS-II} software \cite{toby_gsas-ii:_2013}. In this paper, we use atomic coordinates based on an orthorhombic unit cell (unless otherwise noted) with $b < a < c$ (i.e., $a \approx$ 6.28 \AA, $b \approx 3.496$ \AA, and $c \approx$ 14.07 \AA).

\section{Results and Discussion}

\begin{figure}[h]
\begin{center}
\includegraphics[width=8.6cm]
{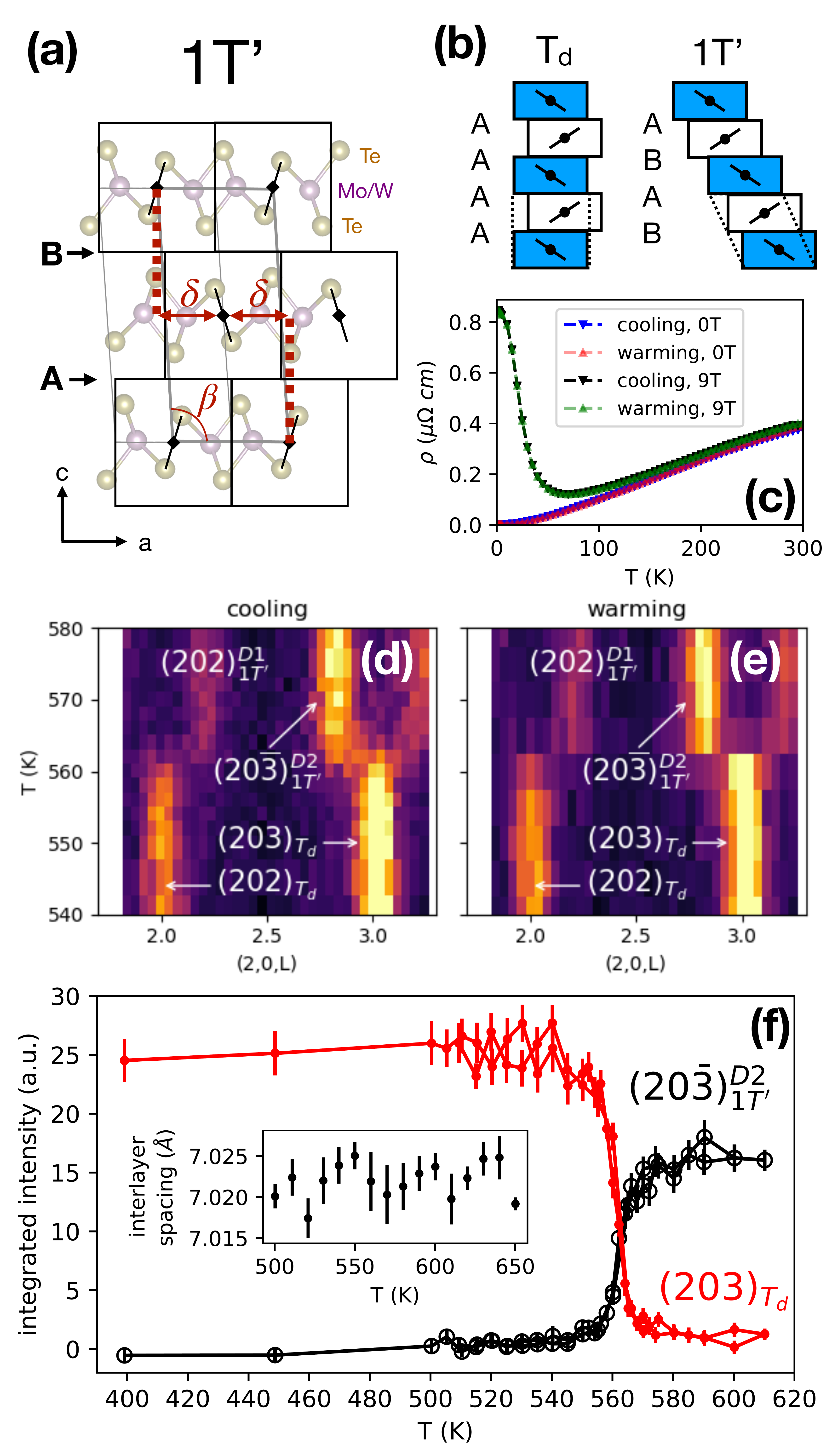}
\end{center}
\caption{(a) The crystal structure of 1T$^{\prime}$-Mo$_{1-x}$W$_{x}$Te$_{2}$ projected in the $a$-$c$ plane. (b) Stacking sequences for the T$_{d}$ and 1T$^{\prime}$ phases of WTe$_{2}$. 
(c) Temperature and field dependence of resistivity in WTe$_{2}$, for current along the $b$-direction and $H \parallel$ $c$. The relative error of each data point is $\sim$0.001.
(d,e) Scans of neutron scattering intensity along (2,0,L) collected on a single crystal of WTe$_{2}$ on cooling and warming. The Bragg peak labelled D1 and D2 refer to the two 1T$^{\prime}$ twins. 
(f) Intensity as a function of temperature of ($(203)_{T_{d}}$ and $(20\bar{3})_{1T^{\prime}}$, obtained from fits of scans along $(2,0,L)$. 
(inset of (f)) The temperature dependence of the interlayer spacing, obtained from fits to longitudinal scans along $(004)$.
}
\label{fig:fig1}
\end{figure}

Shown in Figs.\ \ref{fig:fig1}(d,e) are intensity maps which combine elastic neutron scattering scans along the $(2,0,L)$ at a sequence of temperatures on warming from 510 to 610 K, then cooling. A clear T$_{d}$--1T$^{\prime}$ transition can be seen from changes in the Bragg peaks, which occur without the diffuse scattering seen in MoTe$_{2}$ \cite{tao_appearance_2019}. At low temperatures, the $(202)_{T_{d}}$ and $(203)_{T_{d}}$ Bragg peaks are observed. On warming, a structural phase transition into the 1T$^{\prime}$ phase is observed at $\sim$565 K, followed by 1T$^{\prime}$ phase peaks appearing near $L \approx 2.2$ and $2.8$. The calculated volume fractions of the 1T$^{\prime}$ twins are around 48\% and 52\%. Unlike the appearance of the T$_{d}^{*}$ phase in MoTe$_{2}$, there is no intermediate phase present in the transition in WTe$_{2}$. 

In Fig.\ \ref{fig:fig1}(f), the intensities of the $(203)_{T_{d}}$ and $(20\bar{3})_{1T^{\prime}}^{D2}$ peaks, obtained from fits to scans along $(2,0,L)$, are plotted as a function of temperature on warming and cooling through the hysteresis loop. The transition in WTe$_{2}$ is quite sharp (mostly complete within $\sim$10 K) with very little hysteresis, as seen from the overlap of the warming and cooling curves. In contrast, MoTe$_{2}$ has a broader transition of tens of Kelvins, with a lingering hysteresis in the resistivity that can persist hundreds of Kelvins away from the transition region \cite{tao_appearance_2019}; even the T$_{d}$$\rightarrow$T$_{d}^{*}$$\rightarrow$T$_{d}$ loop in MoTe$_{2}$, which proceeds much more sharply than the transition between T$_{d}$ and 1T$^{\prime}$, still has a hysteresis of $\sim$5 K \cite{tao_appearance_2019}. Although structural phase transitions are often accompanied by anomalies in the lattice constants, no change in the interlayer spacing was observed in the inset of Fig.\ \ref{fig:fig1}(f), in contrast to the abrupt changes seen under pressure for the lattice constants \cite{zhou_pressure-induced_2016}. The $a$-axis did not change dramatically either, given the similar intensities of $(2,0,L)$ scans which were performed across the transition without re-alignment. 

\begin{figure}[h]
\begin{center}
\includegraphics[width=8.6cm]
{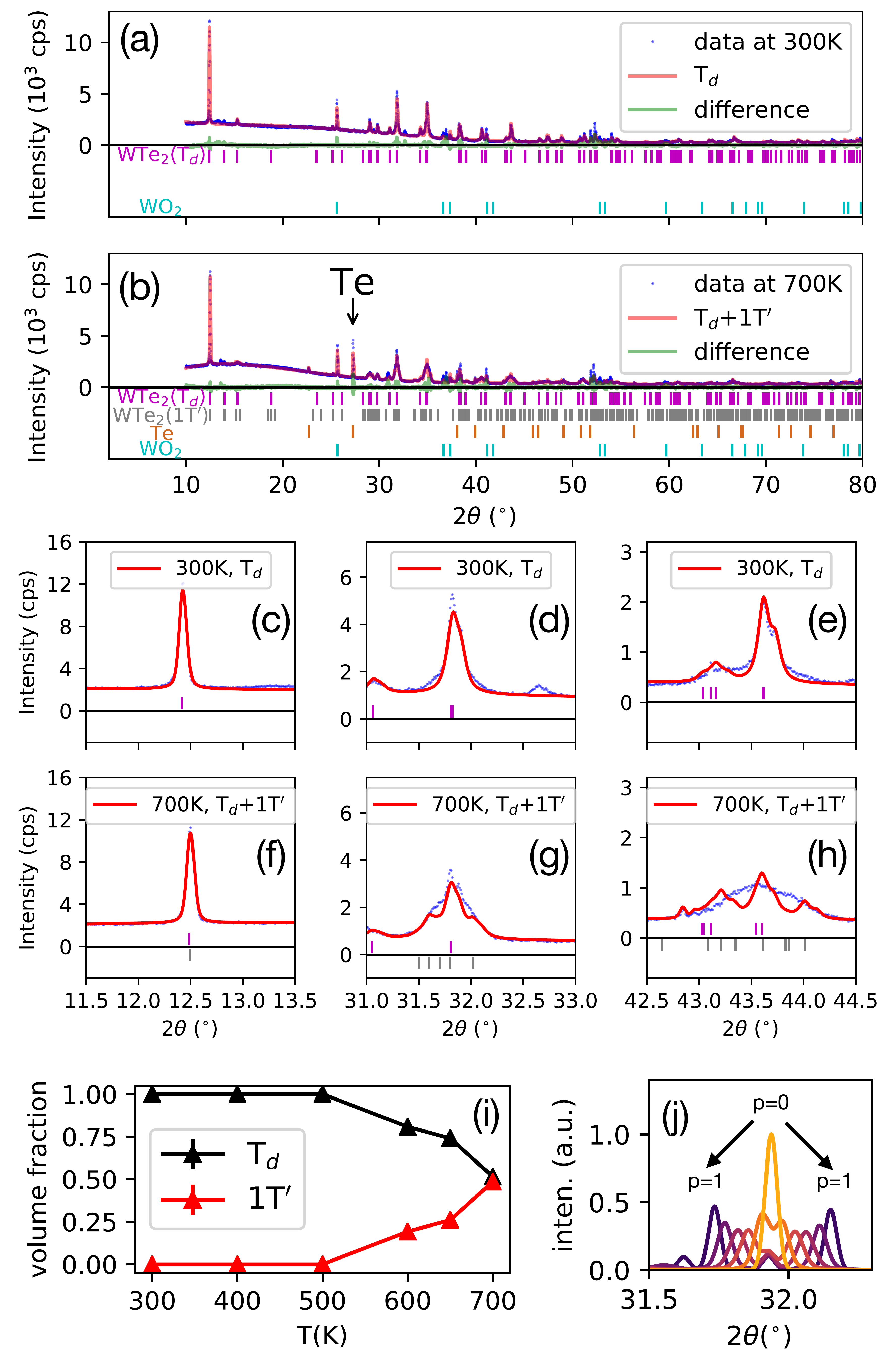}
\end{center}
\caption{(a,b) A plot of the X-ray diffraction pattern compared to the refined model for the average symmetry of powder WTe$_{2}$, collected at 300 K and 700 K on warming. Pure Te Bragg peaks are observed at 700 K. (c-h) Diffraction data plotted in a narrow range (blue dashed lines) for 300 K (c-e) and 700 K (f-h) for the (0,0,2) peak (c,f) and two other peaks. The red curves correspond to the calculated intensity for the T$_{d}$ phase or a T$_{d}$--1T$^{\prime}$ phase coexistence, respectively. (i) The volume fractions of the T$_{d}$ and 1T$^{\prime}$ phases as a function of temperature. (j) Simulated XRD intensities for T$_{d}$ ($p=0$), 1T$^{\prime}$ ($p=1$) and disordered stacking ($0<p<1$), intermediate between T$_{d}$ and 1T$^{\prime}$, for the same regions as (d,g). $p$ is the probability of a randomly swap of “A”  with “B”-type stacking for every other interlayer boundary in the T$_{d}$ AAAA... stacking.}
\label{fig:fig2}
\end{figure}

\begin{figure}[h]
\begin{center}
\includegraphics[width=8.6cm]
{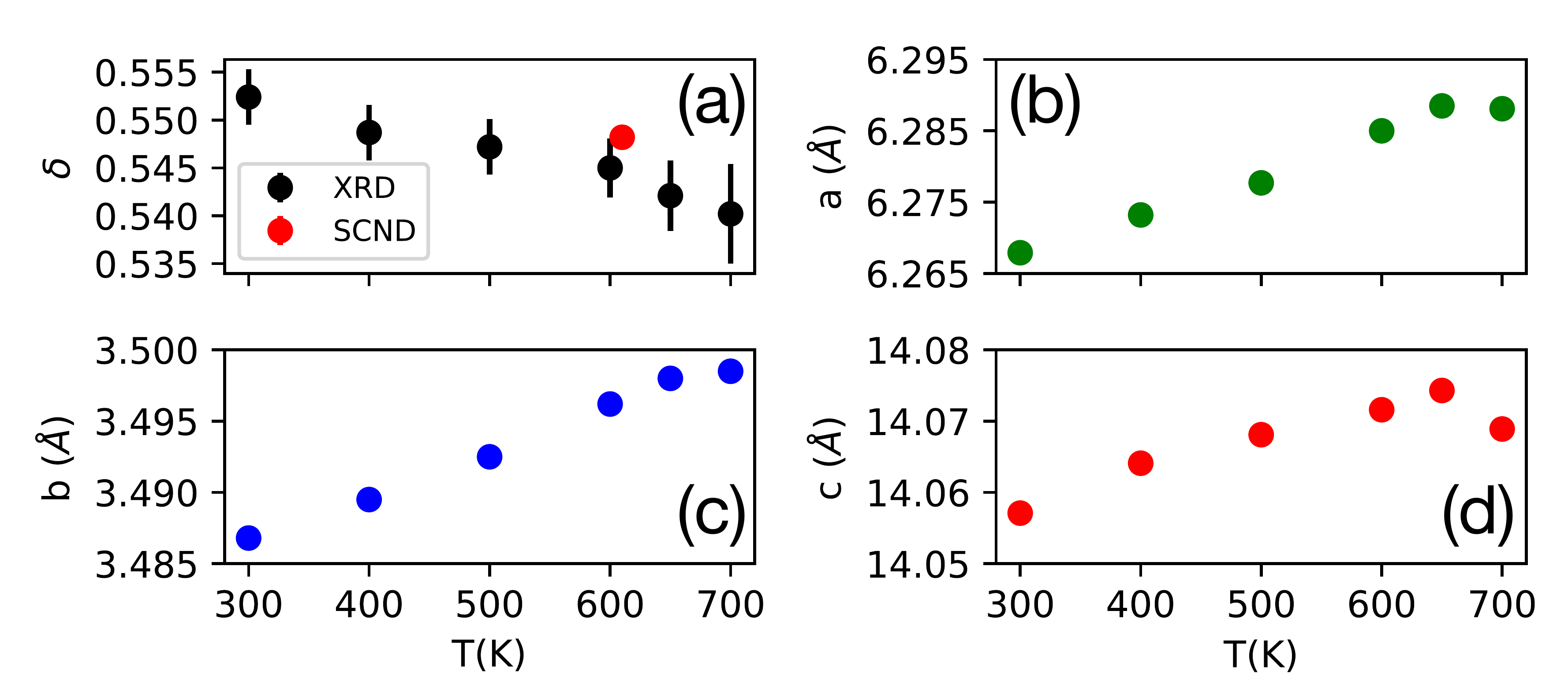}
\end{center}
\caption{(a) The temperature dependence of the $\delta$ parameter of WTe$_{2}$ from powder XRD (black) and single crystal neutron diffraction (SCND) (red) measured on HB1A.
(b-d) The temperature dependence of the lattice constants $a$, $b$, and $c$. 
The error bars for the points in (a-d) are smaller than the symbols except for the XRD $\delta$ points in 3(a).}
\label{fig:fig3}
\end{figure}

In contrast to the clean transition seen in the single crystal, we observe a broad T$_{d}$--1T$^{\prime}$ transition in powder WTe$_{2}$ on warming to 700 K. XRD patterns at 300 and 700 K are shown in Figs.\ \ref{fig:fig2}(a,b).
At 300 K, peaks from the T$_{d}$ phase can be seen, as well as an impurity WO$_2$ phase having weight percent 5.1(1)\%. At 700 K, the pattern can be better fit by a combination of T$_{d}$ and 1T$^{\prime}$, as depicted in Figs.\ \ref{fig:fig2}(c-h). 
The WO$_{2}$ impurity phase was still present at 700 K. Peaks belonging to Te arose, first observable around 600 K, and reaching a weight percent of 8.13(17)\% by 700 K. These Te peaks suggest the decomposition of WTe$_{2}$, though refinement suggested no Te vacancies; a refinement of 700 K data with the occupancies of all Te atoms in T$_{d}$- and 1T$^{\prime}$-WTe$_2$ fixed to a single value yielded a composition of WTe$_{2.016(18)}$. Though decomposition implies that elemental W should be present, no W peaks were seen.

When fitting both T$_{d}$ and 1T$^{\prime}$ to the 700 K data, we first allowed the lattice constants of both phases to vary, then kept the 1T$^{\prime}$ lattice constants fixed to be consistent with those found in the T$_{d}$ refinement; the monoclinic tilting angle $\beta$ was fixed to a value corresponding to the $\delta$ parameter derived from the refined T$_{d}$ phase atomic coordinates. 
With these assumptions, the T$_{d}$ phase can be seen (in Fig.\ \ref{fig:fig2}(d,e) and Fig.\ \ref{fig:fig2}(g,h)) to have its peaks split into those of 1T$^{\prime}$, though the resulting intensity is broader than expected for a simple combination of peak intensities from the two phases. 
If the broadening were due to a spread of lattice constants induced by decomposition, we might expect the $(00L)$ peaks to also be broadened, whereas these peak intensities should not change for changes in stacking (even for disordered stackings) since the $(00L)$ peak intensities only depend on atomic positions along the out-of-plane axis. We indeed see a lack of broadening of the $(002)$ peak in Figures \ref{fig:fig2}(c) and \ref{fig:fig2}(f), suggesting that the change in peak intensity is consistent with changes in stacking. The relative volume fractions for the 1T$^{\prime}$ and T$_{d}$ contributions are shown in Fig.\ \ref{fig:fig2}(i). The transition in the WTe$_{2}$ powder is much broader than in the single crystal, beginning between 500 and 600 K, and steadily increasing up to at least 700 K. 

Disordered stacking likely accounts for the extra intensity between the T$_{d}$ and 1T$^{\prime}$ Bragg peaks. 
To illustrate, we show simulated XRD patterns from disordered stacking sequences progressing from T$_{d}$ to 1T$^{\prime}$ in Fig.\ \ref{fig:fig2}(j). 
While a variety of disordered stackings are conceivable, we used a simple model which was used to analyze diffuse scattering in MoTe$_{2}$ \cite{schneeloch_emergence_2019}. We start from the T$_{d}$ structure with AAAA...\ stacking, then randomly swap ``A'' with ``B''-type stacking with probability $p$ for every other interlayer boundary; thus, $p=0$ corresponds to T$_{d}$, and $p=1$ corresponds to 1T$^{\prime}$ with ABAB...\ stacking. The diffuse scattering can then be obtained from the structure factor of the Bragg peaks from a very large supercell. Fig.\ \ref{fig:fig2}(j) shows simulated patterns for selected $p$ for a 1000-layer supercell. Increasing $p$ results in a steady shift of intensity toward the locations of the 1T$^{\prime}$ peaks. Though the intensity is peaked, even for intermediate $p$, a broader intensity distribution could result from inhomogeneity in the values of $p$, or from a more complex model of stacking disorder. 

An essential parameter characterizing the Mo$_{1-x}$W$_{x}$Te$_{2}$ structure is the $\delta$ parameter, which characterizes in-plane positioning of neighboring layers. 
From the refined coordinates of the T$_{d}$ phase XRD data, we obtained $\delta$ as a function of temperature (Fig.\ \ref{fig:fig3}(a).) The $\delta$ parameter decreases by $\sim$0.007 from 300 to 600 K, which is very similar to the decrease in Mo$_{0.91}$W$_{0.09}$Te$_{2}$ ($\sim$0.006 from 320 to 600 K.) For the 1T$^{\prime}$ phase in the single crystal, we can obtain $\delta$ from the separation between opposite-twin 1T$^{\prime}$ peaks, yielding 0.5482(3) at 610 K (and a monoclinic $\beta$ angle of $92.456(17)^{\circ}$.) This latter value is probably more reliable than those from powder refinement, which may be more insidiously affected by systematic errors due to the indirect nature of obtaining positions from Bragg peak intensities. Nevertheless, a rough agreement for $\delta$ is found between values found from the T$_{d}$-phase powder refinement and from the 1T$^{\prime}$ peak splitting in the single crystal, as seen in Fig.\ \ref{fig:fig3}(a). The refined T$_{d}$-phase lattice parameters are shown in Fig.\ \ref{fig:fig3}(b-d). Aside from a possible anomaly near 700 K, which may be related to the decomposition that results in the Te phase, or to the difficulty in getting uniquely fitted lattice constants in the presence of stacking disorder, we see the expected thermal expansion for $a$, $b$, and $c$. 


Our finding of a T$_{d}$--1T$^{\prime}$ structural phase transition in WTe$_{2}$ suggests that theories of the transition be revisited. Although the relative stability of 1T$^{\prime}$ over T$_{d}$ in MoTe$_{2}$ at higher temperature has been supported by density functional theory calculations  \cite{kim_origins_2017, heikes_mechanical_2018}, WTe$_{2}$ is predicted not to have a transition up to 500 K, and likely much higher \cite{kim_origins_2017}. In MoTe$_{2}$, the preference for 1T$^{\prime}$ at high temperature is attributed to the phonon entropy contribution (with a lack of soft mode behavior noted) \cite{heikes_mechanical_2018}, and more accurate calculations may suggest a similar reason for the existence of 1T$^{\prime}$ in WTe$_{2}$. However, there are two theoretical obstacles. First, there is the inherent difficulty in reliably calculating the very small differences in free energy between phases like 1T$^{\prime}$ and T$_{d}$. Second, beyond the relative stability of 1T$^{\prime}$ and T$_{d}$, to our knowledge no theoretical attempts have been made to explain the details of the transition, including the existence/absence of a hysteresis, presence of T$_{d}^{*}$ on warming, stacking disorder in other parts of the transition, gradual disappearance of stacking disorder on warming/cooling away from the transition, etc.\ \cite{tao_appearance_2019}. Interestingly, the calculations in Ref.\ \cite{kim_origins_2017} show a lack of an energy barrier in WTe$_{2}$ between 1T$^{\prime}$ and T$_{d}$, in contrast to MoTe$_{2}$, which may be related to the lack of hysteresis in WTe$_{2}$ but not in MoTe$_{2}$. However, other factors, such as increased thermal energy facilitating layer movement, may play a role as well.

The structural trends shown in our data place constraints on theoretical models for the transition. We observed no detectable change in the interlayer spacing across the transition, similar to the negligible change seen in other Mo$_{1-x}$W$_{x}$Te$_{2}$ crystals \cite{schneeloch_evolution_2020}. (Kinks in interlayer spacing vs.\ temperature have been seen in some Mo$_{1-x}$W$_{x}$Te$_{2}$ crystals, but may be due to slight misalignment accompanying the transition \cite{schneeloch_evolution_2020}.) This finding highlights the similarities between the phases, expected since they have nearly identical layers that are positioned relative to neighboring layers in nearly symmetry-equivalent ways. Such similarities may make sufficiently accurate calculations difficult, with subtle effects such as spin-orbit coupling contributing non-negligibly to the layer spacing \cite{kim_origins_2017}. Another structural factor to be considered is the dependence of the $\delta$ parameter on composition and temperature, though these trends are less constraining. Theory already appears to be consistent with the decrease in $\delta$ with W-substitution, with calculated values of $\delta=0.540$ for WTe$_{2}$ vs.\ $\delta=0.564$ for MoTe$_{2}$ (as extracted from calculated 1T$^{\prime}$ lattice constants), and experimental values of 0.552 for our powder T$_{d}$-WTe$_{2}$ data vs.\ $\delta=0.574$ reported for 1T$^{\prime}$-MoTe$_{2}$ \cite{heikes_mechanical_2018} (both at 300 K.) The similarity in the temperature-dependence of WTe$_{2}$ and  Mo$_{0.91}$W$_{0.09}$Te$_{2}$ \cite{schneeloch_evolution_2020} suggests that these compositions have a similar anharmonicity in the interlayer potential, despite the difference in $\delta$.

There are several possible explanations for the broadness of the transition in WTe$_{2}$ powder as compared to single crystals. First, Te vacancies may be responsible, as they have been proposed to broaden the transition in MoTe$_{2-z}$ crystals \cite{cho_te_2017}. We would expect that powder would have more decomposition than a single crystal due to a greater surface area to volume ratio. 
However, XRD refinement of the WTe$_{2}$ powder showed no evidence of Te vacancies.
The inaccuracy in the Te occupancy is likely due to difficulties in fitting Bragg peaks when the disordered stacking is present.
A second possibility is that the transition is broadened in the small crystallites of a powder sample. 
In thin MoTe$_{2}$ crystals (hundreds of nm or less) the transition is known to be broadened or suppressed completely \cite{he_dimensionality-driven_2018, zhong_origin_2018, cao_2018}.
Third, there are likely more defects in powder, induced during sintering or grinding. Defects may frustrate layer sliding, and the presence of grain boundaries and interparticle strain would frustrate the shape change accompanying each grain's orthorhombic-to-monoclinic transition. A better understanding of non-ideal behavior, such as that of powder, may help in realizing the potential of stacking changes to influence properties in quasi-two-dimensional materials. 

\section{Conclusion}

Using elastic neutron scattering on a single crystal and XRD on a powder sample of WTe$_{2}$, we observed a T$_{d}$--1T$^{\prime}$ structural phase transition in the Weyl semimetal WTe$_{2}$ at ambient pressure. In the crystal, the transition occurs at $\sim$565 K without hysteresis, but in the powder, the  transition is broadened and incomplete up to 700 K. 
Our results place constraints on theories of the structural behavior of Mo$_{1-x}$W$_{x}$Te$_{2}$, which thus far have not predicted a transition in WTe$_{2}$.

\emph{Note added.} During the review of this paper, a study reporting a structural phase transition in WTe$_{2}$ at 613 K at ambient pressure was published \cite{Dahal2020}.

\section*{Acknowledgements}

This work has been supported by the Department of Energy, Grant number DE-FG02-01ER45927.  A portion of this research used resources at the High Flux Isotope Reactor, which is DOE Office of Science User Facilities operated by Oak Ridge National Laboratory.

\bibliography{WTe2}

\end{document}